\documentclass[]{ceurart}

\usepackage{amsmath}
\usepackage{amssymb}
\usepackage[export]{adjustbox}
\usepackage{algorithm2e}
\usepackage{booktabs}
\usepackage{balance}
\usepackage{comment}
\usepackage{colortbl}
\usepackage{color,soul}
\usepackage{calc}
\usepackage{enumitem}
\usepackage{float}
\usepackage{graphics}
\usepackage{graphicx}
\usepackage{hyperref}
\usepackage{mathtools}
\usepackage{mdwlist}
\usepackage{multirow}
\usepackage{subcaption}
\usepackage{tabularx}
\usepackage[colorinlistoftodos]{todonotes}
\usepackage{xspace} 
\usepackage{xcolor}
\usepackage{paralist}
\usepackage{times}

\newcommand{\proposed}{APAE}
\definecolor{magicmint}{rgb}{0.67, 0.94, 0.82}
\definecolor{melon}{rgb}{0.99, 0.74, 0.71}
\newcommand{\lmjm}{LMJM}
\newcommand{\lmdr}{LMDir}
\newcommand{\nores}{\cellcolor{lightgray}}
\newcommand\defas{\mathrel{\overset{\makebox[0pt]{\mbox{\normalfont\tiny\sffamily def}}}{=}}}
\newcommand{\para}[1]{\paragraph{\textnormal{\textbf{#1}.}}}
\newcommand{\uls}{\begin{itemize}}
\newcommand{\ule}{\end{itemize}}
\newcommand{\ols}{\begin{enumerate}}
\newcommand{\ole}{\end{enumerate}}

\newcommand{\RQ}[1]{\textbf{RQ{#1}}}

\newcommand{\pwua}[1]{\eta_#1(\mathcal{M})}
\newcolumntype{b}{X}
\newcolumntype{s}{>{\hsize=.25\hsize}X}

\begin{document}

\copyrightyear{2023}
\copyrightclause{Copyright for this paper by its authors.
  Use permitted under Creative Commons License Attribution 4.0
  International (CC BY 4.0).}

\conference{QPP++ 2023: Query Performance Prediction and Its Evaluation in New Tasks, co-located with 45th European Conference on Information Retrieval (ECIR) from the 2nd to the 6th of April 2023 in Dublin, Ireland}

\title{On the Feasibility and Robustness of Pointwise Evaluation of Query Performance Prediction}

\author[1]{Suchana Datta}[%
orcid=0000-0001-9220-6652,
email=suchana.datta@ucdconnect.ie,
]

\author[2]{Debasis Ganguly}[%
orcid=0000-0003-0050-7138,
email=debasis.ganguly@glasgow.ac.uk,
url=https://gdebasis.github.io/,
]

\author[3]{Derek Greene}[%
orcid=0000-0001-8065-5418,
email=derek.greene@ucd.ie,
url=http://derekgreene.com/,
]

\author[4]{Mandar Mitra}[%
orcid=0000-0001-9045-9971,
email=mandar@isical.ac.in,
url=https://www.isical.ac.in/mandar-mitra,
]

\address[1]{University College Dublin, Ireland}
\address[2]{University of Glasgow, UK}
\address[3]{University College Dublin, Ireland}
\address[4]{Indian Statistical Institute, Kolkata, India}

\begin{abstract}
\input Despite the retrieval effectiveness of queries being mutually independent of one another, the evaluation of query performance prediction (QPP) systems has been carried out by measuring rank correlation over an entire set of queries. 
Such a listwise approach has a number of disadvantages, notably that it does not support the common requirement of assessing QPP for individual queries.
In this paper, we propose a pointwise QPP framework that allows us to evaluate the quality of a QPP system for individual queries by measuring the deviations between each prediction versus the corresponding true value, and then aggregating the results over a set of queries.
Our experiments demonstrate that this new approach leads to smaller variances in QPP evaluations across a range of different target metrics and retrieval models.
\end{abstract}

\maketitle

\section{Introduction}
Query performance prediction (QPP) methods have been proposed to automatically estimate the retrieval effectiveness for queries without making use of any true relevance information (e.g. \cite{qpp_croft_cikm06,uef_kurland_sigir10}).
In practice, a QPP method allows us to dynamically adjust the processing steps for a query, depending on its initial performance estimate. 
Although estimating the performance of individual queries independently is a common requirement in many downstream tasks (e.g., adaptive query processing \cite{adaptive_prf}), the standard QPP evaluation methodology adopted by the IR research community has previously involved a \textbf{listwise} approach, rather than a \textbf{pointwise} one. This is despite the fact that the latter represents a more appropriate strategy for use in downstream applications. To elaborate,
a listwise approach operates on a \emph{set of queries} $\mathcal{Q}$ by first converting it into an ordered set as induced by the QPP estimated scores $\phi(Q)\,\forall Q \in \mathcal{Q}$. It then computes a rank correlation measure, such as Kendall's $\tau$, 
between the ground-truth ordering of the queries as induced by their average precision (AP) values \cite{Cummins14} or by any other IR metric, such as nDCG \cite{hamed_neuralqpp}.

A major disadvantage of listwise QPP approaches is that evaluation is conducted in a relative manner, so the performance of one query is measured relative to the others. However, a downstream performance estimate of an individual query also needs to be evaluated independently of the other queries. In contrast, a pointwise approach measures the effectiveness on individual queries, and then, if required, aggregates the results over a complete set. This is analogous to measuring the retrieval effectiveness metric MAP by computing the average precision values for individual queries and then aggregating them.
Pointwise evaluation also allows us to carry out a per-query analysis of a method often leading to useful insights. For instance, Buckley \cite{ria} found that, by performing an extensive per-topic retrieval analysis, they were able to identify queries where most IR systems fail to retrieve relevant documents. However, a listwise evaluation methodology is not conducive to performing this kind of detailed per-query analysis.

Another drawback of listwise methods is that they can be overly sensitive to the configuration setup used for evaluation. The two most important such configurations are: i) the target retrieval evaluation metric that induces a ground-truth ordering over the set of queries; ii) the retrieval model used to obtain the top-$k$ set of documents for QPP estimation. Indeed, variations in these configurations can lead to both large standard deviations in the reported rank correlation measures and significant differences in the relative ranks of various QPP systems \cite{dg22ecir}.
To address the limitations of listwise methods, we propose a new QPP evaluation framework, \textbf{Aggregated Pointwise Absolute Errors} (\textbf{\proposed}), which is shown to not only be consistent with the existing listwise approaches, but also to be more robust to changes in QPP experimental setup.

\section{A Framework for Pointwise QPP Evaluation}
\label{sec:propose}

\paragraph{Correlation with listwise ground-truth}
Before describing our new QPP evaluation framework \proposed, we begin by introducing the required notation. Formally, a QPP estimate is a function of the form $\phi(Q, M_k(Q)) \mapsto \mathbb{R}$, where $M_k(Q)$ is the set of top-$k$ ranked documents retrieved by an IR model $M$ for a query $Q \in \mathcal{Q}$, a benchmark set of queries.

For the purpose of listwise evaluation, for each $Q\in \mathcal{Q}$, we first compute the value of a target IR evaluation metric, $\mu(Q)$ that reflects the quality of the retrieved list $M_k(Q)$. The next step uses these $\mu(Q)$ scores to induce a \textit{ground-truth ranking} of the set $\mathcal{Q}$, or in other words, arrange the queries by their decreasing (or increasing) $\mu(Q)$ values, i.e., 
\begin{equation}
\mathcal{Q}_\mu = \{Q_i \in \mathcal{Q}: \mu(Q_i) > \mu(Q_{i+1}),
\, \forall i=1,\ldots,|\mathcal{Q}|-1\}  \}
\end{equation}
Similarly, the evaluation framework also yields a \emph{predicted ranking} of the queries, where this time the queries are sorted by the QPP estimated scores, i.e.,
\begin{equation}
\mathcal{Q}_\phi = \{Q_i \in \mathcal{Q}: \phi(Q_i) > \phi(Q_{i+1}),
\, \forall i=1,\ldots,|\mathcal{Q}|-1 \} 
\label{qpp_listwise_pred}
\end{equation}
A listwise evaluation framework then computes the rank correlation between these two ordered sets
$\gamma(\mathcal{Q}_\mu, \mathcal{Q}_\phi),\,\,\text{where}\,\, \gamma: \mathbb{R}^{|\mathcal{Q}|}\times\mathbb{R}^{|\mathcal{Q}|} \mapsto [0,1]$ is a correlation measure, such as Kendall's $\tau$.

\paragraph{Individual ground-truth}
In contrast to listwise evaluations, where the ground-truth takes the form of an ordered set of queries, pointwise QPP evaluation involves making $|\mathcal{Q}|$ \textit{independent comparisons}. Each comparison is made between a query $Q$'s predicted QPP score $\phi(Q)$ and its retrieval effectiveness measure $\mu(Q)$, i.e.,
\begin{equation}
\eta(\mathcal{Q}, \mu, \phi) \defas \frac{1}{|\mathcal{Q}|}\sum_{Q \in \mathcal{Q}}\eta(\mu(Q), \phi(Q))
\label{eq:pwcorr}  
\end{equation}
Unlike the rank correlation $\gamma$, 
here $\eta$ is a pointwise correlation function of the form $\eta:\mathbb{R}\times \mathbb{R}\mapsto\mathbb{R}$.
It is often convenient to think of $\eta$ as the inverse of a \emph{distance} function that measures the extent to which a predicted value deviates from the corresponding true value.
In contrast to ground-truth evaluation metrics, most QPP estimates (e.g., NQC, WIG etc.) are not bounded within $[0, 1]$. Therefore, to employ a distance measure, each QPP estimate $\phi(Q)$ must be normalized to the unit interval. Subsequently, $\eta$ can be defined as
$\eta(\mu(Q), \phi(Q)) \defas 1-|\mu(Q) - \phi(Q)/\aleph|$,
where $\aleph$ is a normalization constant which is sufficiently large to ensure that the denominator is positive.

\paragraph{Selecting an IR metric for pointwise QPP evaluation}

In general, an unsupervised QPP estimator will be agnostic with respect to the target IR metric $\mu$. For instance, NQC scores can be seen as being approximations of AP@100 values, but can also be interpreted as approximating any other metric, such as nDCG@20 or P@10. Therefore, a question arises around which metric should be used to compute the individual correlations in Equation \ref{eq:pwcorr}. Of course, the results can differ substantially for different choices of $\mu$, e.g., AP or nDCG. This is also the case for listwise QPP evaluation, as reported in \cite{dg22ecir}. To reduce the effect of such variations, we now propose a simple yet effective solution.

\paragraph{Metric-agnostic pointwise QPP evaluation}
For a set of evaluation functions
$\mu \in \mathcal{M}$ (e.g., $\mathcal{M} = \{\text{AP@100}, \text{nDCG@20},\ldots\}$), we employ an aggregation function to compute the overall pointwise correlation (Equation \ref{eq:pwcorr}) of a QPP estimate with respect to each metric.
Formally,
\begin{equation}
\eta(Q,\mathcal{M},\phi) = \Sigma_{\mu \in \mathcal{M}} 
(1-|\mu(Q) - \phi(Q)/\aleph|), \label{eq:avgpwcorr}
\end{equation}
where $\Sigma$ denotes an aggregation function (it does not indicate summation). In particular, we use the most commonly-used such functions as choices for $\Sigma$: `minimum', `maximum', and `average' -- i.e., $\Sigma \in \{\text{avg}, \text{min}, \text{max}\}$.
Next, we find the average over these values computed for a given set of queries $\mathcal{Q}$, i.e., we substitute $\eta(Q,\mathcal{M},\phi)$ from Equation \ref{eq:avgpwcorr} into the summation of Equation \ref{eq:pwcorr}.

\section{Experiments}
\label{sec:exp}
\begin{table}[!t]
\caption{\small
QPP configurations - (QPP method, IR metric, and models) used to measure variations.}
\centering
\begin{adjustbox}{width=.85\columnwidth}
\begin{tabular}{@{}l@{~~~~}|l@{}}
\toprule
QPP Methods & AvgIDF \cite{survey_preret_qpp}, Clarity \cite{croft_qpp_sigir02}, NQC \cite{kurland_tois12}, WIG \cite{wig_croft_SIGIR07}, UEF(Clarity), UEF(NQC), UEF(WIG) \cite{uef_kurland_sigir10} \\
\midrule
IR Metrics & AP@100, nDCG@100, P@10, Recall@100 \\
\midrule
IR Models & \lmjm~($\lambda=0.6$), \lmdr~($\mu=1000$), BM25 $(k,b)=(0.7,0.3)$ \\
\bottomrule
\end{tabular}
\end{adjustbox}

\label{tab:models_and_metrics}
\end{table}

A QPP experiment context \cite{dg22ecir} involves three configuration choices: i) the \textbf{QPP method} itself that is used to predict the relative performance of queries; ii) the \textbf{IR metric} that is used to obtain a ground-truth ordering of the query performances as measured on a set of top-$k$ ($k=100$ in our experiments) documents retrieved by iii) a specific \textbf{IR model}. Table \ref{tab:models_and_metrics} summarizes the IR models and metrics used in our experiments, along with the relevant hyper-parameter values. The objective of our experiments is to investigate the following two key research questions:

\begin{compactitem}
    \item \RQ{1}: Does \proposed~\textit{agree} with the standard listwise correlation metrics?
    \item \RQ{2}: How \textit{robust} is \proposed~with respect to changes in the QPP experiment context?
\end{compactitem}
\begin{table}[!b]
\centering
\caption{
\footnotesize
The correlation of our proposed pointwise evaluation metric \proposed~with the standard listwise metrics - Pearson's $r$, Spearman's $\rho$, Kendall's $\tau$ and sARE. The rank correlations between each pair of QPP system ranks (evaluated with a listwise measure and a pointwise measure) were computed with Kendall's $\tau$.
The high values indicate that the pointwise measurement can effectively \textit{substitute} a standard list-based measure, since they lead to a fairly similar relative ordering between the effectiveness of different QPP methods.
}
\begin{adjustbox}{width=0.75\columnwidth}
\begin{tabular}{@{}l@{~~}c@{~~}c@{~~}c@{~~}c@{~~~~}c@{~~}c@{~~}c@{~~}c@{~~~~}c@{~~}c@{~~}c@{~~}c@{}}
\toprule
&\multicolumn{4}{c}{ $\pwua{\text{avg}}$}
&\multicolumn{4}{c}{ $\pwua{\text{min}}$}
&\multicolumn{4}{c}{ $\pwua{\text{max}}$}
\\

\cmidrule(r){2-5}
\cmidrule(r){6-9}
\cmidrule(r){10-13}

& \multicolumn{1}{c}{$r$} 
& \multicolumn{1}{c}{$\rho$} 
& \multicolumn{1}{c}{$\tau$}
& \multicolumn{1}{c}{$\text{sARE}$}
& \multicolumn{1}{c}{$r$} 
& \multicolumn{1}{c}{$\rho$} 
& \multicolumn{1}{c}{$\tau$}
& \multicolumn{1}{c}{$\text{sARE}$}
& \multicolumn{1}{c}{$r$} 
& \multicolumn{1}{c}{$\rho$} 
& \multicolumn{1}{c}{$\tau$}
& \multicolumn{1}{c}{$\text{sARE}$}
\\

\midrule
BM25 
& 0.810 & 0.810 & 0.905 & 0.887 
& 0.778 & 0.778 & 0.794 & 0.813 
& 0.802 & 0.810 & 0.794 & 0.794 \\

LMDir\;\;\;
& \textbf{0.905} & \textbf{0.810} & \textbf{0.905} & \textbf{0.887} 
& 0.778 & 0.794 & 0.794 & 0.810 
& 0.769 & 0.782 & 0.794 & 0.796 \\

LMJM 
& 0.810 & 0.810 & 0.810 & 0.846 
& 0.794 & 0.794 & 0.782 & 0.786 
& 0.794 & 0.769 & 0.810 & 0.846 \\
\bottomrule
\end{tabular}
\end{adjustbox}
\label{tab:stable}
\end{table}

An affirmative answer to \RQ{1} would indicate that our proposed metric \proposed~is \textit{consistent} with existing metrics used for QPP evaluation, while an affirmative answer to \RQ{2} would suggest that \proposed~is preferable to existing methods due to its higher stability with respect to different experimental settings.

\begin{table}[!th]
\caption{
\footnotesize
Stability of the proposed pointwise QPP metric APAE with respect to listwise approach, across different pairs of IR metrics and IR models. Red cells indicate the lowest value in each group, while the lowest values along each column are bold-faced.
\label{tab:sdres}
}
\vskip 0.5em
\begin{subtable}[t]{.48\linewidth}
\centering
\begin{adjustbox}{width=\textwidth}
\begin{tabular}{@{}l@{~~}l@{~~}c@{~~}c@{~~}c@{~~}c@{~~}c@{~~}c@{}}

\toprule

Model & Metric & 
AP@100 & 
R@10 & R@100 & 
nDCG@10 & nDCG@100 \\


\midrule

\lmjm & \multirow{3}{*}{AP@10} & 
0.497 & 
0.813 & \cellcolor{melon}0.429 &
0.783 & \cellcolor{melon}0.429 \\

BM25 & & 
0.897 & 
0.722 & 0.722 &
0.793 & 0.793 \\

\lmdr & & 
0.897 & 
0.786 & 0.786 &
0.823 & 0.905 \\

\cmidrule{1-2}

\lmjm & \multirow{3}{*}{AP@100} & \nores & 
\cellcolor{melon}\textbf{0.328} & 0.811 &
0.363 & 0.783 \\

BM25 & & 
\nores & 
0.783 & 0.784 &
0.714 & 0.642 \\

\lmdr & & 
\nores & 
0.823 & 0.901 &
0.834 & 0.789 \\

\cmidrule{1-2}





\lmjm & \multirow{3}{*}{R@10} & 
\nores & 
\nores & 0.624 &
0.893 & \cellcolor{melon}0.503 \\

BM25 & & 
\nores & 
\nores & 0.803 &
0.982 & 0.894 \\

\lmdr & & 
\nores & 
\nores & 0.903 &
0.864 & 0.864 \\

\cmidrule{1-2}

\lmjm & \multirow{3}{*}{R@100} & 
\nores & \nores & 
\nores &
0.852 & 0.804 \\

BM25 & & 
\nores & \nores & 
\nores &
0.786 & 0.890 \\

\lmdr & & 
\nores & \nores & 
\nores &
\cellcolor{melon}0.738 & \cellcolor{melon}0.738 \\

\cmidrule{1-2}





\lmjm & \multirow{3}{*}{nDCG@10} & 
\nores & \nores & 
\nores & \nores& \cellcolor{melon}0.537 \\

BM25 & & 
\nores & \nores & 
\nores &
\nores & 0.904 \\

\lmdr & & 
\nores & \nores & 
\nores &
\nores & 0.868 \\





\bottomrule
\label{tab:metric_tau_tau}
\end{tabular}
\end{adjustbox}
\caption{\footnotesize Correlations between the relative ranks of 7 different QPP systems across different pairs of IR target metrics. QPP systems were evaluated with the baseline listwise metric - Kendall's $\tau$. 
\label{stab:tab1}
}
\end{subtable}%
\quad
\begin{subtable}[t]{.48\linewidth}
\centering
\begin{adjustbox}{width=\textwidth}
\begin{tabular}{@{}l@{~~}l@{~~}c@{~~}c@{~~}c@{~~}c@{~~}c@{}}
\toprule

Model & Metric & 
AP@100 & 
R@10 & R@100 & 
nDCG@10 & nDCG@100 \\


\midrule

\lmjm & \multirow{3}{*}{AP@10} & 
0.904 & 
1.000 & \cellcolor{melon}0.715 &
1.000 & 0.792 \\

BM25 & & 
1.000 & 
1.000 & 1.000 &
1.000 & 1.000 \\

\lmdr & & 
1.000 & 
1.000 & 1.000 &
1.000 & 1.000 \\

\cmidrule{1-2}

\lmjm & \multirow{3}{*}{AP@100} & 
\nores & 
0.905 & 0.811 &
\cellcolor{melon}0.669 & 1.000 \\

BM25 & & 
\nores & 
1.000 & 1.000 &
1.000 & 1.000 \\

\lmdr & & 
\nores & 
1.000 & 1.000 &
1.000 & 1.000 \\





\cmidrule{1-2}

\lmjm & \multirow{3}{*}{R@10} & 
\nores & 
\nores & 0.603 &
0.905 & \cellcolor{melon}\textbf{0.542} \\

BM25 & & 
\nores & 
\nores & 1.000 &
1.000 & 1.000 \\

\lmdr & & 
\nores & 
\nores & 1.000 &
1.000 & 1.000 \\

\cmidrule{1-2}

\lmjm & \multirow{3}{*}{R@100} & 
\nores & \nores & 
\nores &
\cellcolor{melon}0.654 & 1.000 \\

BM25 & & 
\nores & \nores & 
\nores &
1.000 & 1.000 \\

\lmdr & & 
\nores & \nores & 
\nores &
1.000 & 1.000 \\

\cmidrule{1-2}





\lmjm & \multirow{3}{*}{nDCG@10} & 
\nores & \nores & 
\nores & \nores &
\cellcolor{melon}0.649 \\

BM25 & & 
\nores & \nores & 
\nores & \nores &
1.000 \\

\lmdr & & 
\nores & \nores & 
\nores & \nores &
1.000 \\





\bottomrule
\label{tab:metric_apae_tau}
\end{tabular}
\end{adjustbox}
\caption{\footnotesize Similar to Table \ref{stab:tab1}, except QPP performance was evaluated with the pointwise approach \proposed. A comparison with Table \ref{stab:tab1} indicates a better consistency in the relative ranks of QPP systems for variations in the IR metrics.
\label{stab:tab2}
}
\end{subtable}
\begin{subtable}[t]{.48\linewidth}
\centering
\begin{adjustbox}{width=\textwidth}
\begin{tabular}{@{}l@{~~}c@{~~}c@{~~}c@{~~}c@{~~}c@{~~}c@{~~}c@{~~}c@{}}

\toprule

\multirow{2}{*}{Metric} & \multirow{2}{*}{Model} & 
\lmjm &
BM25 & BM25 & 
\lmdr & \lmdr \\

& & ($0.6$) &
($0.7, 0.3$) & ($0.3, 0.7$) &
($500$) & ($1000$) \\

\midrule

AP@100 & & 
0.826 & 
0.904 & 0.819 & 
0.714 & 0.895 \\

nDCG@100 & \lmjm & 
0.780 & 
\cellcolor{melon}0.694 & 0.695 & 
0.759 & 0.759 \\

R@100 & ($0.3$) & 
0.824 & 
0.769 & 0.782 & 
0.904 & 0.904 \\


\cmidrule{1-2}

AP@100 & &  
\nores & 
0.703 & 0.712 & 
0.904 & 0.823 \\

nDCG@100 & \lmjm & 
\nores & 
0.781 & 0.827 & 
0.811 & 0.811 \\

R@100 & ($0.6$) & 
\nores & 
0.813 & 0.725 & 
0.731 & \cellcolor{melon}\textbf{0.675} \\


\cmidrule{1-2}

AP@100 & &
\nores & 
\nores & 0.903 & 
0.785 & 0.785 \\

nDCG@100 & BM25 &
\nores & 
\nores & 0.897 & 
0.786 & 0.786 \\

R@100 & ($0.7, 0.3$) &
\nores & 
\nores & 0.812 & 
\cellcolor{melon}0.752 & 0.779 \\







\cmidrule{1-2}

AP@100 & &  
\nores & 
\nores & \nores & 
0.887 & \cellcolor{melon}0.882 \\

nDCG@100 & BM25 & 
\nores & 
\nores & \nores & 
0.901 & 0.895 \\

R@100 & ($0.3, 0.7$) & 
\nores & 
\nores & \nores & 
0.889 & 0.901 \\







\cmidrule{1-2}

AP@100 & &  
\nores & 
\nores & 
\nores & \nores & 0.901 \\

nDCG@100 & \lmdr & 
\nores & 
\nores & 
\nores & \nores & \cellcolor{melon}0.893 \\

R@100 & ($500$) &  
\nores & 
\nores & 
\nores & \nores & 0.903 \\


\bottomrule
\end{tabular}
\end{adjustbox}
\caption{\footnotesize Here rank correlations between the relative ranks of QPP systems are measured across IR model pairs. As in Table \ref{stab:tab1}, QPP systems were evaluated with $\tau$. The numbers alongside the IR models denote their respective parameters.
\label{stab:tab3}
}
\end{subtable}%
\quad
\begin{subtable}[t]{.48\linewidth}
\centering
\begin{adjustbox}{width=\textwidth}
\begin{tabular}{@{}l@{~~}c@{~~}c@{~~}c@{~~}c@{~~}c@{~~}c@{}}

\toprule

\multirow{2}{*}{Metric} & \multirow{2}{*}{Model} & 
\lmjm &
BM25 & BM25 & 
\lmdr & \lmdr \\

& & ($0.6$) &
($0.7, 0.3$) & ($0.3, 0.7$) &
($500$) & ($1000$) \\

\midrule

AP@100 & & 
1.000 & 
1.000 & 1.000 & 
1.000 & 1.000 \\

nDCG@100 & \lmjm & 
1.000 & 
0.864 & 1.000 & 
\cellcolor{melon}0.843 & 0.864 \\

R@100 & ($0.3$) & 
1.000 & 
0.864 & 1.000 & 
1.000 & 1.000 \\


\cmidrule{1-2}

AP@100 & &  
\nores & 
1.000 & 1.000 & 
1.000 & 1.000 \\

nDCG@100 & \lmjm & 
\nores & 
0.914 & 1.000 & 
\cellcolor{melon}\textbf{0.813} & 0.914 \\

R@100 & ($0.6$) & 
\nores & 
1.000 & 1.000 & 
1.000 & 1.000 \\


\cmidrule{1-2}

AP@100 & &
\nores & 
\nores & 1.000 & 
1.000 & 1.000 \\

nDCG@100 & BM25 &
\nores & 
\nores & 1.000 & 
1.000 & 1.000 \\

R@100 & ($0.7, 0.3$) &
\nores & 
\nores & \cellcolor{melon}0.812 & 
0.905 & 1.000 \\







\cmidrule{1-2}

AP@100 & &  
\nores & 
\nores & \nores & 
1.000 & 1.000 \\

nDCG@100 & BM25 & 
\nores & 
\nores & \nores & 
1.000 & 1.000 \\

R@100 & ($0.3, 0.7$) & 
\nores & 
\nores & \nores & 
1.000 & 1.000 \\







\cmidrule{1-2}

AP@100 & &  
\nores & 
\nores & 
\nores & \nores & 1.000 \\

nDCG@100 & \lmdr & 
\nores & 
\nores &  
\nores & \nores & 1.000 \\

R@100 & ($500$) &  
\nores & 
\nores & 
\nores & \nores & 1.000 \\


\bottomrule
\end{tabular}
\end{adjustbox}
\caption{\footnotesize Unlike Table \ref{stab:tab3}, here the QPP outcomes were evaluated by \proposed~(instead of $\tau$).
\label{stab:tab4}
}
\end{subtable}
\end{table}

We conduct our QPP experiments on the TREC Robust dataset, which consists of $249$ topics. Following the standard practice for QPP experiments \cite{hamed_neuralqpp,query_variants_kurland}, we report results aggregated over a total of 30 randomly chosen equal-sized train-test splits of the data. The training split of each partition was used for tuning the hyper-parameters for the QPP method.

\paragraph{Agreement between listwise and pointwise evaluation}
Firstly, we investigate the consistency of \proposed~with respect to three standard listwise QPP evaluation metrics: Pearson's $r$, Spearman's $\rho$ and Kendall's $\tau$; and a pointwise approach, scaled Absolute Rank Error (sARE) \cite{sare}. Since sARE is an error measure, we measure correlations of \proposed~with $1-\text{sARE}$ measures (which for the sake of simplicity, we refer to as sARE in Table \ref{tab:stable}). We experiment with three different instances of \proposed~obtained by substituting the aggregation functions -- avg, min and max as $\Sigma$ in Equation \ref{eq:avgpwcorr}, denoted respectively as $\pwua{\text{avg}}$, $\pwua{\text{min}}$ and $\pwua{\text{max}}$.

The results presented in Table \ref{tab:stable} answer \RQ{1} in the affirmative. Each reported value here corresponds to the rank correlation (Kendall's $\tau$) between the relative ranks of the QPP systems ordered by their effectiveness as computed via one of the standard metrics (one of $r$, $\rho$, $\tau$ or sARE) and \proposed, i.e., one of $\pwua{\text{avg}}$, $\pwua{\text{min}}$ and $\pwua{\text{max}}$). The high correlation values between the standard listwise and the proposed pointwise metrics show that \proposed~can be used as a substitute for the standard listwise evaluation. Notably, we see that the average aggregate function yields the best results, and hence for the subsequent experiments we use $\pwua{\text{avg}}$ as the pointwise evaluation metric.

\paragraph{Variances in relative effectiveness of QPP methods}

To investigate \RQ{2}, we consider the relative stability of QPP system ranks for variations in QPP contexts (i.e., different IR models and target metrics), comparing both listwise and pointwise approaches (see Table \ref{tab:sdres}). To clarify with an example, if working with three QPP methods, say AvgIDF, NQC, WIG, we observe that $\tau$(NQC) $>$ $\tau$(WIG) $>$ $\tau$(AvgIDF) for LMDir as measured relative to AP@100. We expect to observe a similar ordering for a different choice of the IR model and target IR metric, say BM25 with nDCG@100. As in our previous experiments, here we measure the rank correlations between a total of seven QPP systems (see Table \ref{tab:models_and_metrics}) via Kendall's $\tau$.

\section{Concluding Remarks}
Unlike the standard listwise QPP evaluation mechanism of measuring an overall rank correlation with respect to a reference ranking of the queries (in terms of retrieval effectiveness), we have proposed a pointwise evaluation method that computes the relative difference between a normalized QPP score and a true IR evaluation measure (e.g., AP@100 or nDCG@20). Our experiments demonstrated that the proposed metric exhibits a high correlation with  standard listwise approaches and is more robust to changes in  QPP experimental setup than listwise evaluation measures. Using this metric, it should thus be possible to evaluate the effectiveness of different QPP methods on downstream tasks on a per-query basis.

\para{\textbf{Acknowledgement.}} \small The first and the third authors were supported by the Science Foundation Ireland (SFI) grant number SFI/12/RC/2289\_P2.

\bibliography{refs_reduced}

\end{document}